%%
%% final version 5.10
%%revised, may 2003
%%

\input harvmac.tex
\def\p{\partial}
\def\mod{{\rm mod}}

\def\CP{{\cal P}}
\def\ch{{\rm ch}}
\def\cee{ c}

%% REFS
\lref\ads{ M.F. Atiyah, H. Donnelly, and I.M. Singer,
``Eta invariants, signature defects of cusps, and values
of L-functions,'' Ann. Math. {\bf 118} (1983) 131.}
\lref\abs{M.F. Atiyah, R. Bott, and A. Shapiro,
``Clifford Modules,'' Topology {\bf 3} (1964) 3.}
\lref\ahss{M.F. Atiyah and F. Hirzebruch, ``Vector Bundles
and Homogeneous Spaces,'' Proc. Symp. Pure Math. {\bf 3} (1961) 53.}
\lref\asv{M. F. Atiyah, I. M. Singer, ``The index of
elliptic operators: V'' Ann. Math. {\bf 93} (1971) 139.}
\lref\aps{M. F. Atiyah, V. Patodi, and I. M. Singer,
``Spectral asymmetry and
Riemannian geometry.'' Math. Proc. Camb. Phil. Soc.
{\bf 77} (1975) 43; Math. Proc. Camb. Phil. Soc.
{\bf 78} (1975) 405; Math. Proc. Camb. Phil. Soc.
{\bf 79}(1976) 71.}
\lref\bottu{R. Bott and L. Tu, {\it Differential Forms
In Algebraic Topology}, Springer-Verlag 1982.}
\lref\bredon{G.E. Bredon, {\it Topology and Geometry}, Springer GTM 139}
\lref\dfive{
V.M. Buhstaber, ``Modules of Differentials in The Atiyah-Hirzebruch
Spectral Sequence. I'', Math. USSR Sb. {\bf 7} (1969) 299;
``Modules of Differentials in The Atiyah-Hirzebruch
Spectral Sequence. II'', Math. USSR Sb. {\bf 12} (1970) 59.}
\lref\cheegersimons{J. Cheeger and J. Simons, ``Differential Characters
and Geometric Invariants,'' in {\it Geometry and Topology}, J. Alexander
and J. Harer eds., LNM 1167.}
\lref\dmwi{E. Diaconescu, G. Moore, and E. Witten, ``$E_8$ Gauge Theory,
and a Derivation of $K$-Theory from $M$-Theory,'' hep-th/0005090}

\lref\fw{D.S. Freed and E. Witten, ``Anomalies in String Theory
with $D$-branes,'' hep-th/9907189.}
\lref\fh{D.S. Freed and M. J. Hopkins, ``On Ramond-Ramond
Fields and $K$-Theory,'' hep-th/0002027.}
\lref\harris{B. Harris, ``Differential Characters and the Abel-Jacobi
Map'', in {\it Algebraic $K$-theory: Connections with Geometry and
Topology},
J.F. Jardine and V.P. Snaith eds., Nato ASI Series C: Mathematical
and Physical Sciences  -- Vol. 279, Kluwer Academic Publishers,
1989.}

\lref\horava{P. Horava,`` $M$-Theory as a Holographic Field Theory,''
hep-th/9712130, Phys.Rev. D59 (1999) 046004;
P. Horava and M. Fabinger, ``Casimir Effect Between World-Branes in
Heterotic $M$-Theory,'' hep-th/0002073}
\lref\hw{P. Horava and E. Witten,
``Heterotic and Type I String Dynamics from Eleven Dimensions'',
Nucl.Phys. {\bf B460} (1996) 506, hep-th/9510209;
``Eleven-Dimensional Supergravity on a Manifold with Boundary'',
Nucl.Phys. {\bf B475} (1996) 94, hep-th/9603142.}
\lref\milstash{J.W. Milnor and J.D.
Stasheff, {\it Characteristic Classes}, Princeton Univ. Press, 1974.}
\lref\nicolai{H. Nicolai, ``$d=11$ supergravity with local $SO(16)$
invariance,'' Phys. Lett. {\bf 187B}(1987)316; S. Melosch and H. Nicolai,
``New Canonical Variables for $d=11$ supergravity,'' hep-th/9709227}
\lref\topten{E. Witten, ``Topological Tools In Ten-Dimensional Physics'',
in {\it Unified String Theories}, 1985 Santa Barbara
Proceedings, M. Green and D. Gross, eds. World Scientific 1986.}
\lref\vwoneloop{C. Vafa and E. Witten, ``A One Loop Test of
String Duality'',  Nucl.Phys. {\bf B447} (1995) 261, hep-th/9505053.}
\lref\fourflux{E. Witten, ``On Flux Quantization in $M$-Theory
and the Effective Action,'' hep-th/9609122; Journal of
Geometry and Physics, {\bf 22} (1997) 1.}
\lref\imwitten{E. Witten, ``Five-Brane Effective Action In $M$-Theory'',
J. Geom. Phys. {\bf 22} (1997) 103, hep-th/9610234.}
\lref\adscft{E. Witten, ``AdS/CFT
Correspondence and Topological Field Theory'',\hfill\break
JHEP {\bf 9812} (1998) 012, hep-th/98120112.}
\lref\baryons{E. Witten, ``Baryons and Branes in Anti de Sitter
Space, JHEP {\bf 9807} (1998) 006, hep-th/9805112.}
\lref\duality{E. Witten, ``Duality Relations among
Topological Effects in String Theory,'' hep-th/9912086.}
\lref\kapustin{A. Kapustin, ``$D$-branes in a Topologically
Nontrivial B-field,'' hep-th/9909089.}
\lref\landweber{P. Landweber and R. Stong, ``A Bilinear Form for Spin
Manifolds'',
Trans. Amer. Math. Soc. {\bf 300} (1987) 625.}

\lref\massey{W.S. Massey, ``On the Stiefel-Whitney Classes of a
Manifold II'', Proc. AMS {\bf 13} (1962) 938.}
\lref\spingeometry{H.B. Lawson and M.-L. Michelsohn,
{\it Spin Geometry}, Princeton 1989.}
\lref\minmoore{R. Minasian and G. Moore,
``$K$-theory and Ramond-Ramond charge,''
hep-th/9710230; JHEP 9711 (1997) 002.}
\lref\selfduality{G. Moore and E. Witten, ``Self-duality, RR fields
and $K$-Theory,'' hep-th/9912279.}
\lref\stong{R. Stong, ``Calculation of
$\Omega_{11}^{spin}(K({\bf Z},4))$'' in
{\it Unified String Theories}, 1985 Santa Barbara
Proceedings, M. Green and D. Gross, eds. World Scientific 1986.}
\lref\szabo{K. Olsen and R.J. Szabo,
``Constructing $D$-Branes from $K$-Theory,''
hep-th/9907140.  }
\lref\thom{R. Thom, ``Quelques Propri\'et\'es Globales Des
Vari\'et\'es Diff\'erentiables'', Comment. Math. Helv. {\bf 28}
(1954) 17.}
\lref\ght{M. B. Green, C. M. Hull, and P. K. Townsend, ``$D$b-Brane
Wess-Zumino Actions, $T$-Duality, and The Cosmological Constant,'' Phys.
Lett. {\bf B382} (1996) 65, hep-th/9604119.}
\lref\bouwknegt{P. Bouwknegt and V. Mathai,  ``$D$-Branes, $B$ Fields
and Twisted $K$-Theory,'' JHEP {\bf 0003:007,2000}, hep-th/0002023.
}
\lref\as{M. F. Atiyah and G. Segal, to appear.}
\lref\wittenk{E. Witten, ``$D$-Branes And $K$-Theory,''
 hep-th/9810188;  JHEP 9812 (1998) 019.}
%%

%macros
%
\def\tilde{\widetilde}
\def\bar{\overline}

\def\Z{{\bf Z}}

\def\S{{\bf S}}
\def\R{{\bf R}}

\font\zfont = cmss10 %scaled \magstep1
\font\litfont = cmr6

\def\bigone{\hbox{1\kern -.23em {\rm l}}}
\def\ZZ{\hbox{\zfont Z\kern-.4emZ}}
\def\half{{\litfont {1 \over 2}}}

\def\CM{{\cal M}}
\def\Re{{\rm Re ~}}

\def\p{\partial}

\def\R{{\bf R}}

\def\mod{{\rm mod}}

\Title{\vbox{\baselineskip12pt\hbox{hep-th/0005091}
\hbox{IASSNS-HEP-00/38}}}
{\vbox{\centerline{A Derivation of  }
\medskip
\centerline{ $K$-Theory from
$M$-Theory   }}}
\centerline{Duiliu-Emanuel Diaconescu}
\smallskip
\centerline{\it School of Natural Sciences, Institute for Advanced Study}
\centerline{\it Olden Lane, Princeton, NJ 08540, USA}
\smallskip
\centerline{Gregory Moore}
\smallskip
\centerline{\it Department of Physics, Rutgers University}
\centerline{\it Piscataway, New Jersey, 08855-0849}
\smallskip
\centerline{Edward Witten$^*$}
\smallskip
\centerline{\it Dept. of Physics, Caltech, Pasadena, CA 91125}
\smallskip\centerline{\it and CIT-USC Center for Theoretical
Physics, USC, Los Angeles CA}
\medskip
%\baselineskip 18pt

%skip

\noindent
We show how some aspects of the $K$-theory classification
of RR fluxes follow from a careful analysis of the phase
of the $M$-theory action.
This is a shortened and simplified  companion
paper to ``$E_8$ Gauge Theory, and a Derivation of $K$-Theory
from $M$-Theory.''

\vskip 2.5cm
\noindent
$^*$ On leave from Institute for Advanced Study, Princeton, NJ 08540.
\Date{May 3, 2000}
%text of paper

%\draft

\newsec{Introduction}

In the past few years, we have learned that $D$-brane
charges should be thought of in the framework of
$K$-theory \refs{\minmoore,\wittenk,\szabo}. More recently, it has been
realized that the topological classification of
RR fluxes in Type II string theory is also
$K$-theoretic \refs{\duality,\selfduality,\fh}.
Other developments of the past few years,
such as  Matrix Theory, and the AdS/CFT
correspondence, have
shown that $D$-branes play an important role in
the search for a more fundamental formulation
of $M$-theory. It is natural, therefore, to
ask how the $K$-theoretic formulation
of RR charges and fluxes can be formulated
in terms of $M$-theory.

In hindsight, the $K$-theoretic interpretation
of RR fluxes and charges is almost inevitable, given the
existence of Chan-Paton vector bundles on $D$-branes.
But $M$-branes do not carry Chan-Paton bundles or vector fields,
so  the $M$-theoretic
origin of $K$-theory is not manifest.
In this letter, we will outline how some aspects of the
$K$-theoretic formulation of RR charges and fluxes
can in fact be derived from $M$-theory. In brief,
when $M$-theory is formulated on an eleven-manifold
of the form $Y=S^1 \times X$,
we can derive what might be called ``integral
equations of motion'' for the topological class
of the four-form flux of $M$-theory.
These equations state that (on-shell) the four-form
flux of $M$-theory is in fact $K$-theoretic, in
a sense we will make more precise below.
 In order
to keep technical complications to a minimum,
we will make several simplifying assumptions on the
topology of $X$. Further details, without
the simplifying assumptions, can be found in
\dmwi. 

Let $X$ be a compact spin 10-manifold
with a fixed Riemannian metric $g_{\mu\nu}$.
Consider Type IIA superstring theory on $X$ with metric $t g_{\mu\nu}$.
We will study the partition function of the
theory in the limit  $g_{s} \rightarrow 0$,
where $g_s$ is the string coupling,
and then  $t \rightarrow \infty$.
We will then consider  $M$-theory on
$Y=X \times S^1$, with metric
\eqn\mmetric{
ds^2_M = g_s^{4/3}(dx^{11})^2 + t g_s^{-2/3} g_{\mu\nu} dx^\mu dx^\nu.
}
In $M$-theory, we will study the partition function in
the limit $t \rightarrow \infty$ and
then $g_s \rightarrow 0$.
Finally, we will show that the leading
terms in the partition function of $M$ theory and Type IIA
theory are in agreement.

One might at first think the agreement between the
two expressions is trivial, since
 eleven-dimensional supergravity reduces on a circle to
ten-dimensional supergravity. However, things are not
so simple because the $K$-theoretic nature of RR
fluxes implies that the sum over RR field-strengths is
{\it not} simply a sum over all harmonic forms that obey conventional
Dirac quantization.
The quantization law is more subtle, and, as we
will see, there are subtle phases in the action
which are not present in the standard treatments of
supergravity Lagrangians.  The goal will be to derive these
subtleties of the Type IIA theory starting from $M$-theory.
In section 6, we will describe three results, of independent 
interest, which are corollaries of our analysis. 

\newsec{The Type IIA Partition Function}

In the weak coupling limit described above, we consider the NS fields
fixed at some classical (not necessarily on-shell) values.
In this background, we will study the one-loop quantum mechanics of
fermions and the free-field quantum mechanics of the RR fields.
The partition function is accordingly
\eqn\zeeia{
Z_{IIA} \sim \exp\left({- {1\over g_s^2} S_{NS}}\right){\Theta_{IIA}\over
\Delta}
}
where $S_{NS}$ is the action for the NS-sector fields,
$\Delta$ is a product of bose and fermi determinants, and $\Theta_{IIA}$,
which is the factor that we will really focus on, is a theta
function which arises from summing over the fluxes of the RR $p$-form fields.

A more complete treatment of the problem would involve integrating
over the moduli of the NS fields -- notably the metric and the $B$-field --
but in the present discussion we hold these fixed and in particular set
$B=0$.

A complete description  of how to construct $\Theta_{IIA}$ can be found in
\refs{\imwitten,\duality,\selfduality}. We consider the full RR
field-strength $G= G_0 + G_2 + \cdots + G_{10}$. This is a sum of real
differential forms on $X$ of even degree. In IIA supergravity,
there is a Bianchi identity $dG=0$, so
$G$ has a characteristic cohomology class that one can regard as an
element of the even degree cohomology
$H^{\rm even}(X;{\bf R})$.  Because $D$-branes exist, $G$ obeys
a Dirac-like quantization condition, but this condition does
not merely state, as one might guess, that the periods of $G/2\pi$ are
integral or in other words that $G/2\pi$ is associated with an
element of the integral cohomology $H^{\rm even}(X;{\bf Z})$.
Instead, from the existence of $D$-branes and their
couplings to RR fields, one can deduce \refs{\duality,\selfduality,\fh}
that
the topological sectors of RR fields are classified by
an element $x\in K(X)$ of the $K$-theory of $X$.  (In Type IIB
one would use $x\in K^1(X)$ in a similar way.)
For given $x$, the cohomology class of $G$ is
\eqn\onshell{
G(x) = \ch(x) \sqrt{\hat A(X)} .
}
This is the $K$-theory version of Dirac quantization.

For our purposes, to construct the partition function $\Theta_{IIA}$,
we will be summing over on shell RR fields.  For each $x\in K(X)$,
 there is a unique harmonic form in the cohomology class of $G(x)$;
we use this in the sum over RR fields.

The second subtlety is that the total field-strength
$G$ should be considered to be self-dual, $G=*G$,
an equation which at the classical level is hardly compatible with \onshell\
for general metrics.
The resolution of this paradox  is that \refs{\imwitten,\duality} one
interprets $G=*G$ as a statement in the quantum theory.
One sums over ``half'' the fluxes, and these  are indeed
quantized by \onshell. More precisely $\Gamma= K(X)/K(X)_{tors}$
(where $K(X)_{tors}$ is the torsion subgroup)
is a lattice with an integral symplectic form given by
the index of the Dirac operator coupled to $x\otimes \bar y$.
If we let $I(z)$ denote the index of the Dirac operator coupled to a bundle
$z$, and we use the Atiyah-Singer formula for the index, then the
symplectic form $\omega(x,y)$ is defined by
\eqn\sympform{
\omega(x,y) = I(x\otimes \bar y) = \int_X \ch(x\otimes \bar y) \hat A.
}
Poincar\'e duality in $K$-theory implies that this form
is {\it unimodular}. One sums over ``half the fluxes''
by summing  over fluxes associated with $K$-theory
classes in a maximal Lagrangian sublattice
$\Gamma_1 \subset \Gamma$. This sum is the theta function
for the quantum self-dual RR field.

To sketch in somewhat more detail the definition of the theta
function,  we use the symplectic structure
\sympform\ to
give the torus $\CP_K(X)= (K(X)\otimes \R)/\Gamma$
the structure of a compact
phase space. Moreover, this phase space has a metric
\eqn\metric{
\parallel x \parallel^2 = \int G(x) \wedge * G(x) .
}
There is therefore a unique translation invariant complex
structure $J$ on $\CP_K(X)$ such that the metric \metric\ is
of type $(1,1)$. Explicitly, $G_{2p}(Jx) = (-1)^{p+1}*(G_{10-2p}(x))$.
Coherent state quantization
with respect to this complex structure leads to
a unique quantum state, since the symplectic volume of
$\CP_K(X)$  is one. $\Theta_{IIA}$ is the  wavefunction
of this quantum state.  To write it
more explicitly, we now choose a complementary Lagrangian
sublattice $\Gamma_2$ so that $\Gamma = \Gamma_1 \oplus \Gamma_2$.
The lattice vectors in $\Gamma_1, \Gamma_2$ define
``$a$-cycles'' and ``$b$-cycles'' in $\CP_K(X)$, and with
respect to this decomposition we have a
period matrix $\tau$, which is  a quadratic form on
$\Gamma_1\otimes \R$ with positive imaginary part.
Finally, we must define the characteristics of the
theta function. This is the subtlest part of the
quantization procedure. We introduce a function
$\Omega: \Gamma \rightarrow \Z_2$ such that
\eqn\cocycle{
\Omega(x+y) = \Omega(x) \Omega(y)e^{i \pi \omega(x,y)} .
}
Then, we may define the characteristics $\theta\in \Gamma_1,
\phi\in \Gamma_2$ by
\eqn\characteristics{
\eqalign{
\Omega(x) = (-1)^{\omega(x,\phi)} ~{\rm for}~ x\in \Gamma_1\qquad
&
\qquad
\Omega(x)  = (-1)^{\omega(x,\theta)} ~{\rm for}~ x\in \Gamma_2. \cr}
}
For an explanation of why the $\Omega$ function is needed and
the rationale for the definition of the characteristics, see
\refs{\imwitten,\duality}.
Finally, we may write the explicit formula for the theta function:
\eqn\thetiia{
\Theta_{IIA} = e^{-i \pi \Re(\tau(\theta/2))}
\sum_{x\in \Gamma_1} e^{i \pi \tau(x+\half\theta)} \Omega(x).
}
%
%(The characteristic $\theta$ is only defined as an element of
%$ \Gamma_1/2\Gamma_1$, and the prefactor in \thetiia\
%guarantees that $\Theta_{IIA}$ is independent of the representative
%in    $\Gamma_1$.)
%

It remains to identify $\Omega$. There is a (presumably unique)
$T$-duality invariant choice \refs{\imwitten,\duality}   given in terms
of the mod two index of Atiyah and Singer \asv. If $V$ is a real vector
bundle  on $X$, then we define $q(V)$ to be the number, modulo two, 
 of chiral zero modes of the Dirac operator
$\Dsl_V$ coupled to $V$. For $X$ of dimension
$8k+2$, $q(V)$ is a topological invariant.  The definition of
\duality\ is:
\eqn\modtwo{
\Omega(x) = (-1)^{q(x\otimes \bar x)}.
}
It is often useful to  regard $x$ as the
charge of a $D$-brane in Type IIB theory. Then, by a
Born-Oppenheimer argument \selfduality, $q(x\otimes \bar x)$ counts
the number modulo two of fermion zero modes in the Ramond sector
for open strings with boundary conditions defined by $x$ at each end.
This makes the $T$-duality invariance of $\Omega$
manifest.

A few facts about the mod two index will prove useful below.
In general, the mod two
index is not just the reduction modulo two of an ordinary index (which is,
after all, simply zero in 10 dimensions for  $V$  real). It is true
that if the real bundle $V$ can (after complexification)
 be written as $V=x \oplus \bar x$ where $x\in K(X)$
and $\bar x$ is the complex conjugate of $x$, then $q(V)$
equals the mod two reduction of $ I(x)$.
This fact is used \duality\ in showing that $\Omega$
satisfies the cocycle relation \cocycle.
%I think it is a bit misleading to claim the following
%Although there is
%no cohomological formula for the mod two index \asv, when one restricts
%to a Lagrangian sublattice of $K(X)$ there is such a formula,
%given by the characteristics \characteristics.

There are many different choices of sublattices $\Gamma_1$.
Up to an overall normalization, different choices
lead to different descriptions of
the same partition function. In the problem
discussed in this letter, there is a very natural choice of
polarization. To motivate it, consider the behavior of the
kinetic energy
of a non-self-dual field $G$ as we scale the metric $g_{\mu\nu}
\to t g_{\mu\nu}$. The kinetic energies scale as:
\eqn\kinetic{
 t^5 \parallel G_0 \parallel^2
+t^3 \parallel G_2 \parallel^2
+t \parallel G_4 \parallel^2
+t^{-1} \parallel G_6 \parallel^2
+t^{-3}\parallel G_8 \parallel^2
+
t^{-5} \parallel G_{10} \parallel^2.
}
We would like to choose a polarization so that only positive
powers of $t$ show up in the exponential. Otherwise the
sum over fluxes becomes less and less convergent as
$t\rightarrow \infty$, and the terms in the sum do not
accurately reflect the long-distance physics. We will have
only positive powers of $t$ if we take
 $\Gamma_2 $ to be the set of $K$-theory classes $x$
with $\ch_0(x) = \ch_1(x) = \ch_2(x)=0$, and then take
$\Gamma_1$ to be a complementary Lagrangian sublattice.
Working through the definitions of the quantization procedure
one finds:
\eqn\ourpol{\eqalign{
\Theta_{IIA} =  e^{-i \pi \Re(\tau(\theta/2))}
\sum_{x\in \Gamma_1}  & e^{-\pi (t^5 \parallel G_0 \parallel^2
+t^3 \parallel G_2 \parallel^2
+t \parallel G_4 \parallel^2 ) }  \cr &
e^{i \pi \int( G_0 G_{10}-G_2 G_8 + G_4 G_6)} \Omega(x) \cr}
}
where $G$ is understood to be given by  \onshell\ evaluated for
$x+\half \theta$.
It might look at first sight like this is just the standard recipe
for computing the RR partition function as a sum over fluxes of $G_0$,
$G_2$, and $G_4$, with the higher RR fields eliminated using self-duality.
However, the allowed values of the $G_0,G_2$, and $G_4$ fluxes
differ from what one would conventionally guess.  Moreover,
in addition to a factor from the standard kinetic energy of
$G_0$, $G_2$, and $G_4$, the action contains nonstandard
phase factors.  The factor $e^{i \pi \int (G_0 G_{10}-G_2 G_8 + G_4 G_6)}$
(which arises by computing the real part of the $\tau$ function of the
lattice) is, after imposing \onshell\
(which constrains the $G_{2p}$ of $p>2$ in terms of those of $p\leq 2$)
a $120^{th}$ root of unity that is given by a complicated cohomological
formula
and is not part of the standard supergravity formalism.  The sign
factor $\Omega(x)$ is not given by any cohomological formula.

As we have stressed in the introduction, we want to focus on the
behavior for  $t \rightarrow \infty$. The dominant contributions
come from   $K$-theory classes $x\in\Gamma_1$ such that $G_0(x) = G_2(x)=0$.
A glance at \onshell\ shows that these are classes of virtual dimension
zero such that $c_1(x) = 0$. Denoting by   $\Gamma_1'$
the sublattice of such classes,
 the leading term in the partition function
may be simplified to
\eqn\leadingkay{
\Theta_{IIA} =  e^{-i \pi \Re(\tau(\theta/2))}
\sum_{x\in \Gamma_1'} e^{-\pi t \parallel G_4 \parallel^2 }
e^{i \pi \int  G_4 G_6} \Omega(x) .
}
It is also important to include $G_2$ for a more complete comparison
to $M$-theory ($G_0$ has no known origin in $M$-theory, at least for general
backgrounds), but for simplicity,
in the present letter we consider only $G_4$.

\newsec{The $M$-Theory Partition Function}

The partition function for $M$-theory in
the large volume limit  is given
by
\eqn\zeem{
Z_{M} \sim \exp\biggl( -\int_Y \sqrt{g} \CR \biggr) {1\over \Delta_{M}}
\Theta_M
}
where the leading term is the Einstein action of a fixed Riemannian
metric \mmetric\  on a spin 11-manifold $Y$,
 $\Delta_M$ are one-loop determinants
(which we take to be  positive, absorbing the sign in $\Theta_M$),
and $\Theta_M$ is the sum over the classical on-shell configurations
of the $C$-field.

As in Type IIA, there is a subtle quantization law on $G=dC$ as well
as a subtle phase-factor in the path integral
 \fourflux.
The topological quantization of 4-form field-strengths is given
by choosing any element $a\in H^4(Y;\Z)$ and taking $G(a)$ to be 
a certain de Rham representative satisfying
\eqn\geea{
{G(a)\over 2\pi}  = a - \half \lambda .
}
Here $\lambda$ is the degree four class represented at the level of
differential forms by
$-{\tr R\wedge R\over 16\pi^2}$.
It is an integral class on a spin manifold and
 satisfies $2\lambda = p_1$. The contribution of
a field $C$ in the  topological 
sector $a$  to the   $M$-theory partition function is
\eqn\mtheta{
  e^{-\parallel G(a) \parallel^2}
\Omega_M(C),
}
where $G(a)$ is the on-shell field configuration. 
%
%In order to 
%make an explicit comparison to the IIA theta function, we will
%take $Y = X \times S^1$ and evaluate the sum over $a\in H^4(X,{\bf Z})$.
%For such fields, the on-shell configurations consist of harmonic forms
%$G(a)$ satisfying \geea.\foot{In fact, the supergravity equation is
%$d *G = G\wedge G$ so that for generic $a-\half \lambda$,
%$G$ is {\it not} harmonic. However, on
%$Y= X \times S^1$, with $a$ pulled back from
%$X$, this equation is enforced by the $B$ field
%of IIA theory, and in this paper  we are holding $B$
%fixed and zero. }
%

The phase factor $\Omega_M(C)$ is a globally well-defined
version of the familiar supergravity interaction
$\sim \int_Y C G G + \cdots $.
Since $G(a)$ is a nontrivial cohomology class, $C$ is not globally
well-defined as a three-form, and the proper formulation of the phase is
tricky
\fourflux.
We first find a 12-manifold $Z$ such that
$\p Z = Y$ and $a$ extends to $\tilde a\in H^4(Z;\Z)$. The existence of
the
pair $(Z,\tilde a)$ is highly nontrivial, but guaranteed by a result of
Stong \stong. We then define the phase by
\eqn\extend{
\Omega_M(C) =\epsilon \exp \biggl[
2\pi i \int_Z \left({1\over 6} \bigl( \tilde a - \half \lambda)^3 +
 (\tilde a - \half \lambda){\lambda^2 - p_2\over 48 }\right)
 \biggr] ,
}
where $\epsilon$ is the sign of the Pfaffian of the gravitino operator.
In a topologically trivial situation we may
identify $\tilde a - \half \lambda = \tilde G = d \tilde C$
and apply Stokes' theorem to make contact with the more standard
supergravity expressions.

The expression \extend\  is not manifestly well-defined 
since the choice of $Z$ is not unique. It was shown in 
\fourflux\ that this difficulty is most elegantly eliminated 
by using $E_8$ index theory. We will also find the connection 
to $E_8$ useful below. Therefore, let us recall that on 
 manifolds $\CM$ of dimension less than 16,
cohomology classes $b\in H^4(\CM;\Z)$ are in 1-1 correspondence with
topological
classes of $E_8$ vector bundles $V(b)$ on $\CM$ \topten.
We hence consider the $E_8$ bundle $V(\tilde a)$ on $Z$ in the
adjoint representation  and choose a
connection $A$ on $V(\tilde a)$ such that on $Y$
\eqn\ceetoa{
C = {1\over 16 \pi^2} {1\over 30} {\Tr}_{\bf 248}(A d A + {2 \over 3 } A^3)
+
{1 \over 32\pi^2} {\Tr} (\omega d \omega + {2 \over 3} \omega^3).
}
In other words, we interpret $C$ as a Chern-Simons three form
of $E_8$ gauge theory plus gravity.
(It is not obvious that given a $C$-field a corresponding gauge
field $A$ always exists. A  slightly longer argument can be made if
a connection $A$ making \ceetoa\ hold on the nose does not exist.)
We then can evaluate the phase in terms of the $\eta$ invariants of the
Dirac operator $\Dsl_{V(a)}$ and the gauge-fixed Rarita-Schwinger operator
 $ D_{RS} = \Dsl_{TX-3\CO}$ on $Y$. Here
$\CO$ is the trivial real line bundle.
% The
%second operator leads to   purely
%gravitational contributions related to the Rarita-Schwinger field, but is
%not the result of the RS path integral.
Using the APS index theorem
\aps, and the fact that the index of $\Dsl_{V(a)}$ is even in 12
dimensions, we can rewrite the phase as
\eqn\etainvts{
\Omega_M(C)
= \exp\Biggl[ 2\pi i \biggl( {\eta(\Dsl_{V(a)}) + h(\Dsl_{V(a)}) \over 4}
 +   {\eta(D_{RS}) + h(D_{RS}) \over 8}
 \biggr) \Biggr]. }
Here $h$ denotes the number of zero modes of the operator
in question on $Y$.  (The sign $\epsilon$ in $\Omega_M$
is absent  as it cancels against a term that
comes from the APS theorem.)

In order to compare to IIA superstrings, we will now restrict attention
to $Y= X \times \S^1$.
We restrict to fields invariant under rotations of $\S^1$, and, since
we have taken $B=0$ on the Type IIA side, we assume that the $C$-field
is a pullback from $X$.  Under these conditions, $\Omega_M(C)$ is a
topological
invariant that
depends only on the characteristic class $a$, and not on $C$, so we will
denote it as $\Omega_M(a)$.  Moreover, the $\eta$ invariants vanish for
$X\times \S^1$ because of the reflection symmetry of $\S^1$, and only the
contributions from the number $h$ of
zero modes survive.  The $a$-dependent factor is then simply
\eqn\defa{\Omega_M(a)=\exp[i\pi h(\Dsl_{V(a)})/2] .}

Using the standard relation between the radius $R$
of $S^1$ and the IIA string coupling we find:
\eqn\leadingm{
\Theta_M = \sum_{a\in H^4(X;\Z) } e^{-\parallel G( a)\parallel^2} \Omega_M(
a) +
\CO(e^{-1/g_s^2}).
}
where $G(a)$ is the harmonic form in the cohomology class $a-\half \lambda$
and the corrections correspond to field-strengths which have an index tangent to
the
$M$-theory circle or are not invariant under rotations of the circle.
\foot{In fact, the supergravity equation is $\kappa d*G = \half G\wedge G + 
(\lambda^2 - p_2)/48$ for a suitable constant $\kappa$. The components of 
this equation which are pulled back from $X$ are enforced by subsequent 
integration over the $B$-field, which is held fixed (and zero) in this letter. 
The difference from the harmonic form is cohomologically trivial.} 

\newsec{$K$-Theory vs. Cohomology}

We would now like to compare \leadingkay\ to \leadingm. These are, {\it a
priori}, rather different expressions.
One involves a sum over a certain part of $K(X)$ that is related
to $H^4(X;\Z)$  and the other involves
a sum over $H^4(X;\Z)$.
If we included the other RR fields, then on the Type IIA side, we would
be summing over the lattice $\Gamma_1\subset K(X)$, and on the
$M$-theory side we would be summing over degree two and degree
four cohomology
classes of $X$.

In general, $K(X)$ and $H^{\rm even}(X;\Z)$ (the sum of the even degree integral
cohomology groups of $X$) are closely related groups.  If one tensors
with the real numbers, they become isomorphic (by the map that takes
an element of $K(X)$ to its Chern character).
However the integral structures (which determine the Dirac quantization
conditions) are different,
and the torsion subgroups can be very different.
For example,
$H^{\rm even}(RP^{2n+1};\Z) \cong \Z \oplus \Z_2^n$ while
$K(RP^{2n+1};\Z) \cong \Z \oplus\Z_{2^n}$.
  We now describe with
some more precision the relation of $K$ and $H^{\rm even}$ at the integral
level. The reader will find more detail in \dmwi.

Let us first describe how integral cohomology arises from
a $K$-theory class $x$.
For every $K$-theory class $x$, there is a smallest integer $i$ such
that $x$ can be represented as the class of a $(2i-1)$-brane, wrapped
on a $2i$-dimensional submanifold $Q$ of $X$.  The Poincar\'e dual of
$Q$ is a $(10-2i)$-dimensional cohomology class $\alpha$ associated with
$x$.

The map from a $K$-theory class $x$ to an associated cohomology
class $\alpha(x)$ is the first step in a systematic procedure,
known as the Atiyah-Hirzebruch spectral sequence,
for comparing $K$-theory to cohomology.
This map is relevant to us, because the Type IIA formula \leadingkay\
involves a sum over $K$-theory classes $x$ for which $\alpha(x)$
is an element of $H^4(X;\Z)$ (modulo those for which $\alpha(x)$ is
of degree six or higher), while the $M$-theory formula \leadingm\
is a sum over the characteristic class $a\in H^4(X;\Z)$.
We will compare the $M$-theory sum over $a$ to the Type IIA sum over
$\alpha(x)$.

Is it the case that every $a\in H^4(X;\Z)$ is $\alpha(x)$ for
some $x\in K(X)$?  The answer to this question is ``no.''
In ten dimensions, a necessary and sufficient condition for $x$ to exist
is that
\eqn\juny{Sq^3(a)=0,} where $Sq^3$ is a certain cohomology operation,
known as a Steenrod square.  If $x$ exists, we call it a ``$K$-theory lift''
of $a$.  Such an $x$ has virtual dimension zero, $c_1(x)=0$, and $c_2(x)=-a$.

An introduction to the Steenrod squares $Sq^i$
is given in \dmwi. In brief, if $a\in H^k(X;\Z)$ then $Sq^3(a)\in
H^{k+3}(X;\Z)$
may be defined as follows. Let $Q(a)$ be a submanifold that is Poincar\'e
dual
to
$a$ in $X$. Then the normal bundle $N(Q)$ of $Q$ has integral
characteristic
classes $W_i(N(Q))\in H^i(Q;\Z)$ for $i$ odd. These can be pushed into a
tubular neighborhood of $Q$,  allowing us to define
\eqn\defsteensq{Sq^3(a)
=W_3(N(Q))\cup a.}
Similarly, one defines the mod two Steenrod squares for all $i$
(not necessarily odd) as follows.  For
$\bar a\in H^k(X;\Z/2)$, we set $Sq^i(\bar a) = w_i(N(Q)) \cup \bar a\in
H^{k+i}(X;\Z/2)$.
The Steenrod squares obey many identities; the ones we need are
as follows (for sketches of proofs see \dmwi).
 First, the integral and mod two
squares are related. In fact,
 $Sq^3(a) =0 $ if and only if $Sq^2  (a) $ has an integral lift,
that is, if and only if there is an integral class $b$ whose mod
two reduction is $Sq^2(a)$.
Second, it is possible to ``integrate by parts'' with
Steenrod squares. That is, for any $a,b$,
$\int_X a\cup Sq^2b = \int_X Sq^2a \cup b$.
Closely related to the criterion \juny\ for a $K$-theory lift to exist
is the fact that if $x$ is a $K$-theory class with $c_1(x)=0$,
then the higher Chern classes of $x$ obey
\eqn\nustuff{c_3(x)= Sq^2c_2(x)~{\rm mod}~2.}
Finally, $Sq^3 Sq^3=0$, so we may take its cohomology, 
namely, the kernel of $Sq^3$ acting on $H^{\rm even}(X,\Z)$ 
modulo the image of $Sq^3$ acting on $H^{\rm odd}(X,\Z)$. 
In our situation, the cohomology of $Sq^3$ is, essentially, 
$K(X)$.

While the equation $Sq^3(a)=0$ might seem somewhat exotic, it is a close
cousin of a condition that has
already appeared in the physics literature on $D$-branes.
In particular, if we think of $x$ as determining the $D$-brane
{\it charge} (in IIB string theory) of a brane wrapped on $Q$,
 then cancellation of worldsheet global
anomalies implies that $W_3(N(Q))=0$ \fw.  Thus, by \defsteensq\
it follows that $Sq^3(a)=0$ if $a$ is Poincar\'e dual to $Q$.

Let us now apply these remarks to study \leadingkay. We would
like to convert the sum over the sublattice $\Gamma_1'$ in
$K$-theory to a sum over cohomology elements, namely to a sum
over classes $a\in H^4(X;\Z)$ such that $a$ has a $K$-theory lift.
At this point we run into an apparent difficulty. A
$K$-theory lift of $a$, if it exists,  is not unique because
given one lift $x$ it is always possible to add a class $y$
to $x$ where $y$ is any element of the lattice $\Gamma_2$ introduced
earlier (thus $G_{2p}(y)=0$ for $p\leq 2$).
The quantities
$G_6(x)$ and $\Omega(x)$ in \leadingkay\ definitely depend on the
choice of $K$-theory lift, but the product
\eqn\prodphs{
e^{i \pi \int  G_4(x+\half \theta) G_6(x+\half \theta)}
\Omega(x)
}
does not. This can be demonstrated using the
facts noted above. Using \onshell, \characteristics,
and integration by
parts, we first rewrite \prodphs\ as
\eqn\prdphseii{
\exp[-i {\pi \over 4} \int c_2(\theta) c_3(\theta)]
\exp\bigl[i {\pi \over 2} \int (c_2(x) + c_2(\theta))c_3(x)\bigr]
\Omega(x).
}
If we change the $K$-theory
lift by $ x \to x+y$ with $y$ as above then
using the cocycle condition \cocycle, the definition
\characteristics\ of
$\theta$, and  the index theorem one can
verify that \prdphseii\ is unchanged.  At this point, we have shown
that the Type IIA partition function can be written as a sum over
$G_4$ fluxes (along with $G_2$ and $G_0$ if one chooses to include them),
as one would naively expect, but with subtle shifts in the Dirac 
quantization
condition and an exotic sign factor in the sum over fluxes.

Finally,
it remains to translate the characteristic $\theta$ into cohomology.
Again, it is useful to regard $x$ as the $D$-brane charge
of  a brane in IIB theory wrapping some worldvolume $Q$ of smallest
possible dimension.
 As we noted above, the mod two index $q(x\otimes \bar x)$
is given by the number mod two of the fermion zero modes of
a singly-wrapped brane  on $Q$. The classes $x\in \Gamma_2$
correspond to $D(-1)$, $D1$ and $D3$ instantons
in $X$. In the first two cases, the number of zero modes is easily
seen to be even. On the other hand, for a $D3$ instanton,
the number of fermion zero modes is given by the index theorem
to be $\int_{Q} \lambda ~{\rm \mod}~ 2$. We conclude that
$\ch(\theta) = - \lambda + \cdots$. Thus, we can
simplify \leadingkay\ to
\eqn\leadingkayp{
\Theta_{IIA} =
\sum_{a\in H^4(X;\Z): ~Sq^3(a)=0}
e^{-\pi t \parallel a  - \half \lambda \parallel^2  }
e^{i {\pi \over 2} \int (c_2(x(a)) + \lambda)c_3(x(a))}
\Omega(x(a))}
where $x(a)$ is any $K$-theory lift of $a$.
We have now
expressed the $K$-theory sum in terms of
cohomology.  It is now time to
re-examine the $M$-theory sum \leadingm.

\newsec{The Integral Equation of Motion in $M$-Theory}

In the previous section we reduced the $K$-theory partition function
to a sum over a subgroup of $H^4(X;\Z)$.   This subgroup is
of finite index, since for any $a$,  $Sq^3(a)$ is of
order 2, and hence $Sq^3(2a)=0$. By contrast, the $M$-theory
partition function is a sum over the full group $H^4(X;\Z)$.
To show that the two expressions for the partition functions
agree, we will argue that  the
 $M$-theory phase $\Omega_M(a)$ leads to  an
``integral equation of motion'' $Sq^3(a)=0$ on the topological
sectors in  $M$-theory.

In this letter we will, for simplicity,
show agreement of \leadingkayp\ and \leadingm\ under the
assumption that $\Omega_M(\cee)=1$ for torsion $\cee$, and that
$Sq^3(\cee)=0$ for all torsion elements $\cee$.
%
%These assumptions are relaxed in \dmwi, and the analysis
%there reveals a new topological condition on $X$ for
%$M$-theory to be well-defined, namely $W_7(X)=0$. This
%condition is guaranteed by our simplifying assumptions.
%
Suppose $a\in H^4(X;\Z)$ is any class and $\cee \in H^4(X;\Z)_{tors}$.
The kinetic energy   of $G$ in the topological sector $a+\cee$ is
identical to that in the sector $a$ because the
field-strength defined by \geea\ is a real differential form
and hence $G(a+\cee) = G(a)$. Since the torsion subgroup is finite we may  equally
well write \leadingm\ as
\eqn\avtor{
\Theta_M = \sum_{a\in H^4(X;\Z)} e^{-\parallel G(a) \parallel^2}
\Omega^{\rm av}_M(a)
}
with
\eqn\barom{
\Omega^{\rm av}_M(a) = {1\over \vert H^4(X;\Z)_{tors}\vert} \sum_{\cee \in
H^4(X;\Z)_{tors}} \Omega_M(a+\cee) .
}
This is useful because the $M$-theory
phase $\Omega_M(a)$ is not independent of the
shift $a \rightarrow a+c$. Indeed, 
the bundles $V(a+\cee)$ and $V(a)$ are definitely
not isomorphic, and, as we will demonstrate below, 
$\Omega^{\rm av}_M$ is in fact a projection operator.
Under a simplifying topological assumption (described below)
this operator is: 
\eqn\projop{
\eqalign{
\Omega^{\rm av}_M(a) & = 0 \qquad\qquad\qquad ~ {\rm if} \quad Sq^3(a) 
\not=0\cr
& = \Omega_M(a) \qquad\qquad {\rm if} \quad Sq^3(a)  =0. \cr}
}
Moreover, when $Sq^3(a)=0$, so that $a$ has  a $K$-theory lift $x\in K(X)$,
we can compare $M$-theory and $K$-theory phases. We will show that they agree
\eqn\phaseagree{
\Omega_M(a) = \Omega(x)e^{{i \pi \over 2} \int_X (c_2(x) + \lambda)
c_3(x) } .
}

The agreement of   \leadingkayp\ with \leadingm\  immediately follows from
the above pair of results. We will now sketch how they are derived,
  beginning with the proof of \phaseagree. It is here that
the interpretation of the $M$-theory phase in terms of $E_8$
gauge theory is particularly effective. We are interested in
$Y=X \times S^1$ with supersymmetric spin structure on the $M$-theory
circle.  In evaluating \defa, we use the fact that
 a zero mode of $\Dsl_{V(a)}$ is constant along
the $M$-circle so that the phase just depends on the number of
 zero modes on $X$. These may be expressed in terms of the number
of chiral zero modes in 10 dimensions via $h(\Dsl_{\pi^*V(a)}) =
h^+(\Dsl_{V(a)}) +
h^-(\Dsl_{V(a)})= 2 h^+(\Dsl_{V(a)})$. We conclude that the
phase is expressed in terms of a mod two index,
$f(a)=q(V(a))$:
\eqn\eatemodtwo{
\Omega_M(a) =  (-1)^{f(a)}.
}

The next step is to relate the $E_8$ bundle $V(a)$ to a $K$-theory class
$x$. In general, $K$-theory classes are differences of vector bundles
$x=E_1 - E_2$ where the structure group of $E_i$ must be taken to
be $U(N)$ for some large $N$. However, we are working in 10-dimensions,
and this dimension is sufficiently small that {\it all $K$-theory classes
on $X$ with $c_1(x)=0$ can be realized using $SU(5)$ bundles}. The
 reason for this is that  the classification
of bundles on a ten-manifold
depends only on the homotopy groups $\pi_i(SU(N))$ for $i<10$.
(See \topten\ for a description of this approach.) In
10 dimensions, $SU(5)$ is in the stable range: $\pi_i(SU(5))=
\pi_i(SU(\infty))$, $i<10$.
We can therefore take our $K$-theory lift to be $x=E-F$ where
$F$ is
a trivial  rank 5  bundle and $E$ is an $SU(5)$ bundle
with $\ch(E)= 5 + a + \cdots$.
We can construct an $E_8$ bundle $V(a)$ with characteristic class $a$
using the ``embedding'' of $SU(5)\times SU(5)$ in $E_8$, taking the two
$SU(5)$ bundles to be respectively $E$ and $F$.
Using the decomposition of the adjoint representation of $E_8$
under $SU(5)\times SU(5)$ and the fact that $F$ is trivial, one
finds that for mod 2 index theory (throwing away representations
that appear an even number of times), the ${\bf 248}$ is equivalent to
$E\otimes \bar E-{\cal O}+\wedge^2 E+\wedge^2\bar E$, where ${\cal O}$
is a trivial line bundle and $\wedge^2$ denotes the second antisymmetric
product.
Using the properties of the mod 2 index described above we now learn that
\eqn\compareii{
q(V(a)) = q(E \otimes \bar E - \CO )+ I(\Lambda^2(E))
= q(x \otimes \overline{x}) + I(\Lambda^2(E)) + I(E)~{\rm mod}~2.
}
%  We have used the fact that, as $x=E-F$ with $F$ a trivial rank
%5 bundle, $q(x(a)\otimes \overline{x(a)})=q(E\otimes\bar E-{\cal O})
%+I(E)$ mod 2.
The formula \compareii\ leads directly to
\phaseagree.  Indeed,
the first term on the RHS of \compareii\ corresponds to
$\Omega(x)$, while by an easy application of the index theorem,
 the second term is
$\half\int (c_2(x) + \lambda) c_3(x) ~\mod~ 2$.

It remains to show \projop. This is based on an analog of \cocycle\
for the $E_8$ mod two index $f(a)$. Namely, $f$ satisfies the
bilinear identity
\eqn\bilinear{
f(a + a') = f(a) + f(a') + \int_X a\cup Sq^2 a' .
}
 Unfortunately, there does
not appear to be an elementary proof of \bilinear. A proof using ``cobordism
theory'' can be found in section 3.2 of \dmwi.
Granted this, we are now ready to
complete the proof of \projop. The argument simplifies 
considerably if we assume that $f(c)=f_0$ is independent of 
$c$ for {\it torsion } classes $c$. In this case we can   write
\eqn\provepo{
\Omega^{\rm av}_M(a) = (-1)^{f(a)+f_0}
{1\over \vert H^4(X;\Z)_{tors}\vert} \sum_{\cee \in
H^4(X;\Z)_{tors}} e^{ i \pi \int \cee \cup Sq^2 a } .
}
Now, for any $b\in H^6(X;\Z)$ and any torsion $\cee$
it is always true that $\int b \cup \cee = {1\over n} \int b \cup (n\cee) = 0
$.
Using Poincar\'e duality, one can prove the
converse: if $\bar b\in H^6(X;\Z_2)$ satisfies
$\int c\cup  \bar b =0 $ for all $\cee \in H^4(X;\Z)_{tors}$,
then $\bar b$ is the reduction of some integral class.
Therefore  $\Omega^{\rm av}_M(a)$ projects onto the set of
classes $a$ such that $Sq^2 a$ has an integral
lift. This is equivalent
to $Sq^3(a)=0$, i.e., to the statement that $a$ has a
$K$-theory lift $x$. Indeed $Sq^2a$ is the reduction
modulo two of $c_3(x)$.
This completes the proof of \projop, and therefore establishes the
equivalence of \leadingkay\ and \leadingm.

\newsec{Three applications}

The $K$-theory/$E_8$-formalism described above leads 
to three interesting physical effects which we will 
sketch very briefly here. 

First, an easy consequence of \bilinear\ leads to a 
new topological consistency condition on string 
backgrounds. By \bilinear\ we have $f(a + 2 c) = f(a) + f(2c)$, 
and moreover $f(2c) = \int c \cup Sq^2 c$. By a result of 
Stong \stong\ $\int c \cup Sq^2 c= \int c\cup  Sq^2 \lambda$. 
Now, the reasoning below \provepo\ shows that 
$\Omega_M^{\rm av}(a)=0$, and hence $\Theta_M=0$ if 
$Sq^3\lambda\not=0$. In algebraic topology one shows 
that a certain characteristic class $W_7(X)$ of $X$ 
is $Sq^3(\lambda)$. Thus $W_7(X)=0$ is a necessary 
condition for a consistent background. Unfortunately, 
we do not know an intuitive interpretation of this condition.

Second, it turns out that the parity symmetry of 
$M$ theory on $X \times S^1$ (coming from reflection of the $S^1$) is 
anomaly-free,
but this depends on a surprising anomaly cancellation between
bosons and fermions. In IIA theory, 
this symmetry is $(-1)^{F_L}$, and in IIB it is 
related to strong/weak coupling duality. By counting 
fermion zero-modes one can show that the gravitino measure 
$\mu$ transforms under parity as $\mu \rightarrow (-1)^{q(TX)}\mu$. 
The mod-two index $q(TX)$ is nonvanishing for certain 
10-folds, such as $X= T^2 \times {\bf HP}^2$. The fermion 
anomaly is cancelled by the nontrivial transformation law of 
the $G$-flux partition function \leadingm. Indeed, 
parity acts as $G \rightarrow -G$ and by \geea\ 
$a \rightarrow \lambda-a$. Using the bilinear 
identity \bilinear\ and $\int a \cup Sq^2 a = \int a \cup Sq^2 \lambda$, one
finds that $\Theta_M \rightarrow (-1)^{f(\lambda)}\Theta_M$. 
It follows that the total parity 
anomaly is $(-1)^{q(TX)+f(\lambda)}$. On the other hand, 
the fermion measure of the {\it heterotic string} on 
$X$ transforms under $(-1)^F$ by the same factor 
$(-1)^{q(TX)+f(\lambda)}$. It is shown in \topten\ 
that the heterotic string measure is well-defined, 
so we conclude that $q(TX) + f(\lambda)=0 ~\mod ~ 2$, 
and hence that parity is a good symmetry of $M$-theory. 

Our third application concerns the instability of some 
Type IIB branes wrapping homologically nontrivial cycles.
The $K$-theory interpretation of $D$-branes means that $D$-branes cannot
be wrapped on certain cycles; it also means that $D$-branes wrapped on
certain cycles are unstable even though the cycles are nontrivial in
homology.  
 Let $Q$ be a 
cycle Poincar\'e dual to an integral class $a\in H^{\rm even}(X,\Z)$. 
If $Sq^3(a)\not=0$ then, as we have mentioned, we cannot 
wrap a $D$-brane  on $Q$. However, as stressed near 
\nustuff, $K(X)$ is, essentially, the {\it cohomology} of 
$Sq^3$. Thus, if $a$ is ``closed,'' that is
$Sq^3(a)=0$, then $a$ can be lifted to a $K$-theory class $x$,
 but if $a$ is ``exact,'' that is $a= Sq^3(a_0)$ for some $a_0$, 
then one can take $x=0$.
A $D$-brane whose lowest RR charge is given by such an $a$ can in fact be
unstable, even though the class $a$ is nonzero in cohomology.
Annihilation of such a $D$-brane occurs
 via nucleation and subsequent annihilation of $D9-\overline{D9}$
pairs. This follows from the $K$-theoretic interpretation 
\wittenk\ of the work of Sen on brane-antibrane annihilation.

\newsec{Conclusions, Further Results, and Open Problems}

The matching described above and in \dmwi\ 
between the $M$-theory formalism based on $E_8$ and
the Type IIA formalism based on $K$-theory gives considerable added
confidence in both.
In particular, we gain added confidence that not only
$D$-brane charges, but also RR fluxes, should be
classified by $K$-theory. This is an important conceptual change
from the $K$-theoretic classification of $D$-brane
charge; among other things, it suggests that RR fields,
and not just $D$-branes, should somehow be associated with 
vector bundles.

We have focused here on the simplest
case of the computation of \dmwi\ in order to illustrate
some of the central ideas. 
The simplifying topological assumptions
we have made are relaxed in \dmwi.
Also, in  \dmwi\ we extend the computation sketched above to include
 $G_2\not=0$ in
Type IIA theory; in $M$-theory, this corresponds letting
$Y$ be a circle bundle over $X$ with Euler class $ G_2/2\pi$.
After a lengthier analog of the above computation with some additional
ingredients added,
the phases turn out to agree.

One interesting general lesson that emerges from  
 \dmwi\ is that when  one takes torsion
into account there is no direct relation between the flux
$G_4$ in IIA theory and the four-form $G$ in $M$-theory.
They have different underlying integral quantizations,
and  there is no 1-1 correspondence of the terms in
the $M$-theory and the IIA theta functions.
It is only after applying the ``integral equation of
motion,'' $Sq^3(a)=0$ that one can compare results.

As for future directions,
it should be very interesting to
compare the absolute normalization of 
the $M$-theory and Type IIA
partition functions; this depends on
the one-loop determinants as well as some
other overall normalization constants which
arise when $Sq^3$ does not annihilate the torsion.
One would like to extend the computation to
include $D$-brane and $M$-brane instanton effects.
Another, more difficult, open problem concerns
the proper interpretation of nonzero values of
$G_0$. While it  is straightforward  to include the effects
of $G_0$ in the IIA partition function,
comparing the results to $M$-theory presents an
interesting and unsolved challenge.

Our computation
confirms the utility of relating the $C$-field of $M$-theory
to $E_8$ gauge theory as in \fourflux.
Other clues of a possible role of $E_8$ in the formulation of $M$-theory
include
 the possibility of writing eleven-dimensional
supergravity in terms of  gauge fields of a noncompact form of $E_8$
\nicolai, evidence for
 propagating $E_8$ gauge fields  in $M$-theory
on a manifold with boundary
 \hw, and further issues considered in \horava.

Finally, we mention that these considerations lead to an unresolved
question in the case of Type IIB superstring theory.
The problem is to reconcile
the $SL(2,\Z)$ symmetry of this theory
with the $K$-theoretic interpretation of RR charges and
fluxes. Although we have found some nontrivial
partial results relevant to this problem,
the main  puzzle remains unsolved. Nevertheless,  we hope
that the clarification of the relation of $M$-theory
and $K$-theory will play some role in the resolution.

\bigskip
\centerline{\bf Acknowledgments}

We would like to thank M. F. Atiyah, D. Freed,
J. Harvey, M. J. Hopkins,  C. Hull,
P. Landweber, J. Morgan, G. Segal, and
I. M. Singer  for discussions and
explanations.
The work of GM is supported by DOE grant DE-FG02-96ER40949.
The work of EW has been supported in part by NSF Grant
PHY-9513835 and the Caltech Discovery Fund.
The work of DED has been supported by DOE grant DE-FG02-90ER40542.
DED would also like to thank D. Christensen, K. Dasgupta, J. Gomis,
C. Rezk and especially L. Nicolaescu and S. Stolz for useful
discussions.
\bigskip

\listrefs

\bye